\documentstyle[12pt]{article}

\input amssym.tex

\newsymbol\Real 2052
\newsymbol\Comp 2043

\newcommand{\Idd}{/\!\!/}
\newcommand{\Idu}{\backslash \!\!\backslash}

\newcommand{\ket}[1]{\left\vert\,{#1}\,\right>}

\newcommand{\bu}[1]{\left<\,{#1}/\right.}
\newcommand{\kd}[1]{\left./{#1}\,\right>}

\newcommand{\bd}[1]{\left<\,{#1}\backslash\right.}
\newcommand{\ku}[1]{\left.\backslash{#1}\,\right>}

\newcommand{\bk}[2]{\left<{#1}\right\vert\left.{#2}\right>}

\newcommand{\bkdu}[2]{\left<\,{#1}\,\backslash\,{#2}\,\right>}
\newcommand{\bkud}[2]{\left<\,{#1}\,/\,{#2}\,\right>}

\newcommand{\bkuu}[2]{\left<\,{#1}\,/\!\backslash\,{#2}\,\right>}
\newcommand{\bkdd}[2]{\left<\,{#1}\,\backslash\!/\,{#2}\,\right>}

\newcommand{\gu}{/\!\backslash}
\newcommand{\gd}{\backslash\!/}

\newcommand{\idd}{/}
\newcommand{\idu}{\backslash}

\newcommand{\opd}[1]{/{#1}/}
\newcommand{\opu}[1]{\backslash{#1}\backslash}

\newcommand{\lind}{L({\cal K},{\cal K})}
\newcommand{\linu}{L(\hat{\cal K},\hat{\cal K})}

\newtheorem{defin}{Definition}
\newtheorem{theor}{Theorem}

\newenvironment{demo}
{\bgroup\par\smallskip\noindent{\it Proof: }}{\rule{0.5em}{0.5em}
\egroup}

\begin{document}

\title{Generalized ``bra-ket"  formalism}

\author{Ion I. Cot\u aescu.\\ {\small \it  West University of 
Timi\c soara,}\\ {\small \it  V. P\^ arvan Ave. Nr. 4, RO-1900 
Timi\c soara, Romania}}

\date{\today}

\maketitle

\begin{abstract}
The Dirac's bra-ket formalism is generalized to  finite-dimensional vector 
spaces with indefinite metric in a simple mathematical context similar to
thatof the theory of general tensors where, in addition, scalar products are 
introduced with the help of a metric operator. The specific calculation 
rules are given in a suitable intuitive notation. It is shown that the 
proposed bra-ket calculus is appropriate for the general theory 
of basis transformations and finite-dimensional representations of the 
symmetry groups of the metric operators. The presented  application is the 
theory of finite-dimensional representations of the $SL(2,\Comp)$ group with 
invariant scalar products.

Pacs: 02.10.Sp, 02.20.Qs
\end{abstract}
\

\newpage

\section{Introduction}

The bra-ket  calculus invented by Dirac \cite{D} for unitary or Hilbert 
spaces is one of the most coherent, efficient and elegant formalisms. This 
is based on the natural relation among vectors and functionals represented 
by  ket  and respectively  bra vectors. In this way, one obtains simple 
calculation rules which  include the axioms of the scalar product as well as 
the consequences of the Frechet-Riesz theorem. Grace of these qualities, the 
Dirac formalism  offers us the opportunity of working with general 
operator relations, independent on the concrete representations that can 
have very different features in the general case of the rigged Hilbert 
spaces used in quantum theories.

However, in many problems  the vector spaces  can not be organized as unitary, 
Euclidian or Hilbert spaces. We refer especially to the spaces of 
finite-dimensional linear representations of the  non-compact Lie groups 
\cite{G,BR} where the bilinear forms we need to construct invariants are not 
positive definite \cite{W}. In this case one uses the tensor calculus with  
indices in upper and lower positions and bilinear (or inner) forms defined 
with the help of a metric tensor. For unitary,  Euclidian or 
Hilbert spaces the tensor calculus is equivalent with the  Dirac formalism in 
a representation given by an orthonormal basis but for the  spaces with 
indefinite metric \cite{B} we have not yet a suitable Dirac formalism which 
should reproduce all the mechanisms of calculus with covariant and 
contravariant indices. For this reason  we would like to propose here 
a generalization of the Dirac formalism to  vector spaces with indefinite 
metric. We restrict ourselves only to the finite-dimensional case when these 
spaces are called  semi-unitary or semi-Euclidian \cite{ON}.  

The main problem here is the generalization of the mutual bra-ket relation to 
spaces with indefinite  metric. In our opinion, this problem can not be 
solved using only one vector space and its dual space. We mean that 
starting with a space considered as being of covariant vectors (with 
contravariant components) and with its  dual space of contravariant 
functionals, 
we can not relate  covariant vectors with  contravariant functionals in a 
satisfactory bra-ket formalism. This could be achieved only by introducing 
new 
ingredients, namely contravariant vectors and covariant functionals, which 
should allow one to relate vectors and functionals of same kind (covariant or 
contravariant). Therefore, the generalization of the Dirac formalism  requires 
to doubly the number of vector spaces like in the theory of general tensors 
(with simple and dotted indices) where one uses four vector spaces associated 
with the four unequivalent fundamental representations of the group of general 
linear transformations \cite{W}. This means that we have already the framework 
we need. It remains only to correctly define  mutual  bra-ket relations in 
accordance with the mentioned exigency of a coherent bra-ket mechanism.      
We show  that this can be done with the help of two anti-linear 
mappings that conserve the covariance. These will relate the spaces of  
covariant and contravariant ket vectors with their corresponding spaces of bra 
vectors. Moreover, we {\em couple} between themselves the spaces of ket vectors 
as well as those of bra vectors through an isometry compatible with 
the considered anti-linear mappings. In these conditions we can use the operator 
of this isometry as {\em metric operator}. This will replace the metric tensor 
taking over its role in defining scalar products.    

Our objective is to generalize the Dirac bra-ket formalism in this mathematical 
context which combines the framework of the general tensor calculus with the 
theory of bilinear forms given by metric operators \cite{N}. We establish the 
basic calculation rules in suitable notations and  we verify that these lead to 
the common tensor calculus in any representation given by a system of dual 
bases. Actually, we shall see that our formalism recovers all the main results 
of the theory of the general linear transformations but expressed only in terms 
of components with usual covariant or contravariant indices. This is because 
in the bra-ket formalism, where the complex conjugated components appear 
naturally as the components of bra vectors, the artifice of dotted 
indices is no more needed. On the other hand, we show that in 
our approach we have all the technical advantages of the standard bra-ket 
formalism. One of them is that we can work directly with general relations 
involving operators instead of their matrix elements in particular 
representations. Moreover, we can manipulate simultaneously different    
spectral representations of these operators allowing 
us to easily study the theory of finite-dimensional representation of the 
non-compact Lie groups, including those of the symmetry group of the metric 
operator.

The first step is to precise the mathematical framework of our attempt. This 
is presented  in the second section where we define the mutually related  
pairs of  coupled vector spaces which  allows us to introduce the metric 
operator and correct relations among  vectors and functionals. The next 
section is devoted to the specific symbols and notations we propose for 
vectors and operators. In section 4  we construct two kind of compatible 
hermitian forms, called dual forms and scalar products respectively, while 
in section 5 we define the hermitian and Dirac conjugations  for all the 
linear operators we work. Section 6 is devoted to the theory of orthogonal 
projection operators which help us to define  pairs of coupled subspaces with 
hermitian  metric operators. The  matrix representations in systems of dual 
bases are studied in section 7 pointing out the advantages of the orthonormal 
ones. The theory of basis transformations is briefly treated in section 8 
where we give the form of the general linear transformations and we study 
the symmetry transformations that leave invariant the form of the scalar 
product. The example  we propose in section 9 is the theory of 
finite-dimensional representations of the $SL(2, \Comp)$ group with invariant 
scalar products.

\section{Coupled vector spaces}

Our construction is based on a pair of complex vector spaces, 
${\cal V}$ and $\hat{\cal V}$, of {\em covariant} vectors, $x,y,... \in 
{\cal V}$, and 
{\em contravariant} vectors, $\hat x, \hat y,...\in \hat{\cal V}$, respectively. 
The  dual  of ${\cal V}$, denoted by $\overline{\cal V}$, is the space 
of contravariant functionals, $\bar x,...$, while the dual of $\hat{\cal V}$, 
denoted by $\overline{\hat{\cal V}}$, is the space of covariant functionals, 
$\bar{\hat x}...$. The values of these functionals are 
$\bar x(y)$ or $\bar{\hat x}(\hat y)$.  

The basis vectors and the vector components are labeled by Latin indices 
$i,j,k,...$ while the first Latin ones, $a,b,...$ are held for current needs. 
We suppose that all the spaces we work  are finite-dimensional of 
dimension $N$ such that $i,j,...=1,2,...,N$. Furthermore,  we consider 
the covariant basis $\{e_{i}\}\subset {\cal V}$, the contravariant 
basis $\{\hat e^{i}\}\subset \hat{\cal V}$, and the {\em canonical dual} bases, 
from which 
$\{\bar{e}^{i}\}\subset \overline{\cal V}$ is the contravariant one while 
$\{\bar{\hat e_{i}}\}\subset \overline{\hat{\cal V}}$ is the covariant  one. 
Since we assume that these bases satisfy the usual duality conditions, 
\begin{equation}\label{(dual)}
\bar{e}^{i}(e_{j})=\delta^{i}_{j}\,,\qquad   
\bar{\hat e}_{i}(\hat e^{j})=\delta^{j}_{i}\,,
\end{equation}
we  say that these form  a {\em system of dual bases}.

A coherent bra-ket mechanism  requires to {\em mutually relate} among 
themselves the vectors and  functionals. To this end, we introduce the 
anti-linear mappings which conserve the covariance, 
$\phi : {\cal V}\to \overline{\hat{\cal V}}$ and $\hat\phi 
: \hat{\cal V}\to \overline{\cal V}$, defined by       
\begin{equation}\label{(phi)}
\phi[e_{i}]=\bar{\hat e}_{i}\,,\quad 
\hat \phi[\hat e^{i}]= \bar{e}^{i}\,,
\end{equation} 
such that
\begin{equation}\label{(phixy1)}
\phi[x]=(x^{i})^{*}\bar{\hat e}_{i}\,,\quad 
\hat \phi[\hat y]= (\hat y_{i})^{*}\bar{e}^{i}\,,  
\end{equation} 
for any  $x=e_{i}x^{i}\in {\cal V}$ and  $\hat y=\hat e^{i}\hat y_{i} \in 
\hat{\cal V}$.  Then from (\ref{(dual)}) we find   
\begin{equation}\label{(phixy)}
\phi[x](\hat y)=(x^{i})^{*}\hat y_{i}=\left(\hat\phi[\hat y](x)\right)^{*}\,.
\end{equation}
In other respects, equations (\ref{(phixy1)}) show that 
 $\overline{\hat{\cal V}}\sim {\cal V}^{*}$ and
 $\hat{\cal V}\sim \overline{{\cal V}^{*}}$ where  ${\cal V}^{*}$
is the complex conjugate vector space of ${\cal V}$. Therefore, our 
system of mutually related vector spaces is similar to that of the theory of 
general tensors, 
 $({\cal V}, {\cal V}^{*}, \overline{{\cal V}}, \overline{{\cal V}^{*}})$.

Let us consider now the isomorphism  $\eta : {\cal V}\to \hat{\cal V}$,  
defined by the {\em invertible} matrix $|\eta|$ as
\begin{equation}\label{(etae)}
\eta\,e_{i}=\hat e^{j}\eta_{ji}.
\end{equation}
This {\em couples} each covariant vector $x=e_{i}x^{i}$ with the contravariant 
vector $\hat x=\eta\, x =\hat e^{i} x_{i}$ of components $x_{i}= \eta_{ij}
x^{j}$. The corresponding  isomorphism of the dual spaces, $\bar\eta: 
\overline{\cal V}\to \overline{\hat{\cal V}}$, is given by 
\begin{equation}     
\bar{\eta}\,\bar{e}^{k}=(\eta^{-1})^{kj}\bar{\hat e}_{j}
\end{equation}
so that   $\bar\eta\,\bar y(\eta\,x)= y(x)$. 
Moreover, it is not difficult to verify that the isomorphism $\eta$ 
is compatible with the mappings $\phi$ and $\hat\phi$ (closing the diagram) 
only if   
\begin{equation}\label{(comp)}
\phi[x]=\bar\eta\,\hat\phi[\eta\,x], \quad \forall x\in {\cal V}.
\end{equation} 
\begin{theor}
The condition {\rm (\ref{(comp)})} is accomplished if and only if the 
matrix $|\eta|$ is hermitian, i.e. $\eta_{ij}=(\eta_{ji})^{*}$.  
\end{theor}
\begin{demo}
Let us take $x=e_{i}$ and calculate $\bar{\hat e}_{i}=
\bar\eta\,\hat\phi[\eta\,e_{i}]$. 
According to (\ref{(etae)}) and (\ref{(phi)}), this gives
$\bar{\hat e}_{i}=(\eta_{ji})^{*}\bar\eta\,\hat\phi[\hat e^{j}]
=(\eta^{-1})^{jk}(\eta_{ji})^{*}\bar{\hat e}_{k}$ from which we obtain the 
desired  result.
\end{demo}

In what follows we consider only invertible  operators $\eta$ with hermitian 
matrices in a given system of dual bases. They  will be used as the {\em metric 
operators} that define the {\em bilinear forms} of the spaces 
${\cal V}$ and $\hat{\cal V}$,  
\begin{eqnarray}\label{(bilf)}
h(x,y)&=&\phi[x](\eta\,y),\quad ~~~ x,y \in {\cal V},\\ 
\hat h(\hat x,\hat y)&=&\hat\phi[\hat x](\eta^{-1}\,\hat y),\quad  
\hat x,\, \hat y \in \hat{\cal V}.
\end{eqnarray}
A little calculation  giving us the bilinear forms in terms of vector 
components points out that these are {\em hermitian} since 
\begin{equation}\label{(herm)}
h(x,y)=h(y,x)^{*}\,,\quad
\hat h(\hat x,\hat y)=\hat h(\hat y,\hat x)^{*}.
\end{equation}
In addition, we can verify that 
\begin{equation}\label{(izom)}
\hat h(\eta\,x,\, \eta\,y)=h(x,y)
\end{equation}
 which means that the isomorphism $\eta$ is in fact an {\em isometry} 
when $|\eta|=|\eta|^{+}$. 
\begin{defin}
The spaces ${\cal V}$ and $\hat{\cal V}$ isometric through the metric operator 
$\eta$ represent a pair of coupled vector spaces {\rm (cvs)} denoted by 
$({\cal V},\hat{\cal V},\eta)$. The corresponding dual cvs are
$(\overline{\cal V},\overline{\hat{\cal V}},\bar\eta)$. We say that the 
mappings $\phi$ and $\hat\phi$ mutually relate these pairs of cvs.
\end{defin}

\section{Notations}

The  above defined related pairs of cvs represent the appropriate framework 
of our generalized  Dirac formalism. Let us start with the notations of the 
basic elements. 
\begin{defin}
The spaces ${\cal K}\equiv {\cal V}$ and $\hat{\cal K}\equiv \hat{\cal V}$ 
are the spaces of  {\em ket-down}  vectors,  $\kd{~} \in {\cal K}$, or  
of  {\em ket-up}  vectors, $\ku{~}\in \hat{\cal K}$.
\end{defin}    
Thus in our formalism the covariant vectors appear as ket-down vectors while 
the contravariant ones as ket-up vectors. The bra vectors mutually related with 
these ket vectors can be defined with the help of the mappings $\phi$ and 
$\hat \phi$.   
\begin{defin}
The bra vector related to the ket-down vector $\kd{x}$ is the {\em bra-down}  
vector $\bd{x}= \phi[\kd{x}] \in {\cal B}\equiv \overline{\hat{\cal V}}$ 
while the bra vector related to the ket-up vector $\ku{\hat x}$ is the 
{\em bra-up} vector $\bu{\hat x} = \hat\phi[\ku{\hat x}] \in \hat{\cal B}
\equiv \overline{\cal V}$.
\end{defin}
In this manner we have related the  covariant ket vectors of ${\cal K}$ 
with the covariant bra vectors of ${\cal B}$, which are just  the  
covariant functionals defined on the other ket space, $\hat{\cal K}$. 
Similarly, the spaces of contravariant vectors, $\hat{\cal K}$ and 
$\hat{\cal B}$, are also related between themselves even though  the 
contravariant  functionals of $\hat{\cal B}$ are defined on ${\cal K}$. 
Of course, the mutually related ket and bra vectors will be denoted 
systematically with the same symbol as in the usual bra-ket formalism.  

Apparently these crossed bra-ket relations seem to be forced but this is the 
unique way to obtain well-defined hermitian bilinear forms compatible with 
the natural duality. In other respects, our bra-ket relations are correct in 
the sense that the bra vector related with a linear combination of ket 
vectors is the corresponding anti-linear combination of bra vectors. 
Obviously, this is because the mappings $\phi$ and $\hat\phi$ are anti-linear 
(e.g. $\phi[\alpha\kd{x}+\beta\kd{y}]=\alpha^{*}\bd{x}+\beta^{*}\bd{y}$). 
 
Let us consider now the linear operators defined on our cvs. In general, we 
denote by $L({\cal V},{\cal V}')$ the set of the linear operators which map 
${\cal V}$ onto ${\cal V}'$. In our case, the main pieces are the algebras 
$\lind)$ and $\linu$, but we are also interested by the operators which map 
the spaces ${\cal K}$ and $\hat{\cal K}$ to each other. We start with the 
observation that it is natural to denote by  $\Idd\in \lind$ and 
$\Idu\in \linu$ the {\em identity} operators of these algebras since they 
act upon the 
ket-down and respectively ket-up vectors. Moreover, this notation helps us 
to indicate the action of the operators of different kind  by writing them 
between  identity operators. Thus, the operators of $\lind$ can be denoted 
either simply by $A,B,...$ or by  $\opd{A}, \opd{B},...$ in order to avoid  
possible confusions with the operators of the other algebra, 
$\opu{\hat A},\opu{\hat B},...\in \linu$, or with those from 
$L({\cal K},\hat{\cal K})$ or $L(\hat{\cal K},{\cal K})$ that have to be 
delimited by both identity operators in a suitable order. On the other hand, 
the notation we propose has the advantage of indicating the allowed algebraic 
operations. For example, it is clear that the operators 
$\idu A\idd\in  L({\cal K},\hat{\cal K})$ and  
 $\idd B\idu\in  L(\hat{\cal K},{\cal K})$ can be multiplied to each other 
 while their sum does not make sense.  

A special case is that of the metric operators, $\eta\in L({\cal K},
\hat{\cal K})$ and $\eta^{-1}\in L(\hat{\cal K},{\cal K})$, which play the 
central role in our construction. They will be represented by  
\begin{equation} 
\eta\equiv \gd \,,\quad \eta^{-1}\equiv \gu \,,
\end{equation}
in order to obtain the intuitive calculation rules 
\begin{equation}
\gu \gd = \Idd \,,\quad \gd\gu=\Idu\,.
\end{equation}
With these symbols, the pair of cvs of ket vectors can be denoted now by 
$({\cal K}, \hat{\cal K}, \gd)$ while the related pair of bra cvs is    
$(\hat{\cal B},{\cal B},\gu)$. The isometry $\gd : {\cal K}\to \hat{\cal K}$  
couples not only the vectors $\kd{x}$ and $\ku{\hat x}=\ku{\eta x}\equiv
\gd\!\!\kd{x}$, but also couples the   
operators $\opd{A}\in \lind$ and $\opu{\hat A}\in \linu$ through    
\begin{equation}\label{(asop)}
\opu{\hat A}= \gd A \gu \,. 
\end{equation}

\section{Hermitian forms}

Here  we can define two kind of brackets. The first one represents the values 
of  covariant or contravariant functionals.  
We denote by $\bkud{\hat  x}{y}\equiv \hat\phi[\hat x](y)$ the value of the 
contravariant functional $\bu{\hat x}$ calculated for the vector $\kd{y}$, 
and by  $\bkdu{y}{\hat x}\equiv\phi[y](\hat x)$, 
the value of the covariant functional $\bd{y}$ 
calculated for $\ku{\hat x}$.
We  say that the mappings 
$\bkud{~}{~}: \hat{\cal K}\times {\cal K}\to \Comp$ and
$\bkdu{~}{~}: {\cal K}\times \hat{\cal K}\to \Comp$ are {\em  dual 
forms}.   
From (\ref{(phixy)}) it results that the dual forms are {\em hermitian}, i.e.  
\begin{equation}\label{(bk1)}
\bkdu{y}{\hat x}=\bkud{\hat x}{y}^{*}\,, \quad \forall\, 
\kd{y}\in {\cal K},\,\ku{\hat x}\in \hat{\cal K}.
\end{equation}  

The second kind of brackets are just the hermitian bilinear forms defined by 
(\ref{(bilf)}). Since the metric operator is invertible, these bilinear forms 
are nondegenerate and, therefore, can  be called {\em scalar products} 
\cite{ON}. In our new notation the scalar product of ${\cal K}$,
$\bkdd{~}{~}: {\cal K}\times {\cal K}\to \Comp$, has the values
\begin{equation}
\bkdd{x}{y}= 
\bd{x}\!\!\gd\!\!\kd{y}\equiv  h(x,y)\,, \quad \kd{x},\kd{y}\in {\cal K},
\end{equation}
while that of $\hat{\cal K}$, $\bkuu{~}{~}: \hat{\cal K}\times\hat{\cal K}\to 
\Comp$, gives  
\begin{equation}
\bkuu{\hat x}{\hat y}= \bu{\hat x}\!\!\gu\!\!\ku{\hat y}
\equiv \hat h(\hat x,\hat y)\,, \quad \ku{\hat x},\ku{\hat y}\in \hat{\cal K}\,.
\end{equation}
The equations (\ref{(herm)}) which show that these scalar products are hermitian 
take the form 
\begin{equation}
\bkdd{x}{y}=\bkdd{y}{x}^{*}\,,\quad \bkuu{\hat x}{\hat y}=
\bkuu{\hat y}{\hat x}^{*}\,.
\end{equation}\label{(izom1)}
Other useful relations can be written starting with  the coupled vectors 
$\ku{\hat x}=\gd\!\!\kd{x}$ and $\ku{\hat y}=\gd\!\!\kd{y}$. Thus we can find 
equivalences among the values of the dual forms and those of the scalar 
products (e.g. $\bkdu{x}{\hat y}=\bkdd{x}{y}$, $\bkud{\hat x}{y}=\bkuu{\hat x}
{\hat y}$, etc.) or to recover  equation (\ref{(izom)}) giving us the 
isometry of cvs,      
\begin{equation}\label{(xyxy)}
\bkdd{x}{y}=\bkuu{\hat x}{\hat y}\,.
\end{equation}

All these brackets can be imagined as resulting from the traditional  
``juxtaposition" of the Dirac formalism \cite{D}. For example, we can write  
$\bkdu{y}{\hat x}=(\bd{y})(\ku{\hat x})$, $\bkdd{x}{y}=\bd{x} 
(\gd\!\!\kd{y})=(\bd{x}\!\!\gd)\kd{y}$, and so on. Hence the conclusion is that 
we can combine ket and bra vectors of any kind in order to write brackets. If 
the slash-lines are parallel  we obtain a dual form but when these are not parallel, 
leaving an empty angle, then we understand that therein is a metric operator 
giving us a scalar product.   

The orthogonality is defined by the scalar product which play  the same 
role as those  of unitary or Hilbert spaces with the difference that  
here the ``squared norm" $\bkdd{x}{x}$  can take any real 
value, including $0$ even for vectors $\kd{x}\not=0$.  
From (\ref{(xyxy)}) we see that if two ket-down vectors are orthogonal then 
their coupled ket-up vectors are also orthogonal. As mentioned before, the cvs 
are isometric in the sense that for two coupled ket vectors we have  
$\bkdd{x}{x}=\bkuu{\hat x}{\hat x}$. Consequently, the separation of the orbits 
for which this number is positive, negative or zero, can be done simultaneously 
for both cvs \cite{B}.  

\section{Hermitian and Dirac conjugations}

Now we have all the elements for defining the  hermitian 
conjugation that gives us the {\em hermitian adjoint} operators of our 
linear operators. First we consider the operators from 
$L({\cal K},\hat{\cal K})$ and $L(\hat{\cal K},{\cal K})$ and we define:
\begin{defin}
The  hermitian adjoint  operators of $\idu A\idd \in L({\cal K}, \hat{\cal K})$ 
and $\idd B\idu \in L(\hat{\cal K}, {\cal K})$ are 
$\idu A^{+}\idd \in L({\cal K}, \hat{\cal K})$ 
and  $\idd B^{+}\idu \in L(\hat{\cal K}, {\cal K})$ 
which satisfy
\begin{eqnarray}
\bd{x}\!A^{+}\!\kd{y}=\bd{y}\!A\!\kd{x}^{*}\,,\quad \forall\, \kd{x}\,,\kd{y}\in 
{\cal K}\,,\\
\bu{\hat x}\!B^{+}\!\ku{\hat y}=\bu{\hat y}\!B\!\ku{\hat x}^{*}
\,,\quad \forall\, \ku{\hat x}\,,\ku{\hat y}\in \hat{\cal K}\,.
\end{eqnarray}
If $A=A^{+}$ or $B=B^{+}$ we say that these operators are {\em hermitian}. 
\end{defin} 
Obviously, from (\ref{(herm)}) we see that the metric operators are hermitian,  
\begin{equation}
\gd^{+}=\gd\,,\quad \gu^{+}=\gu\,.
\end{equation}

For other operators the situation is more complicated as it results from 
the following definitions. 
\begin{defin}
The  hermitian adjoint  operator of $/A/\in L({\cal K},{\cal K})$ is the 
operator $(\opd{A})^{+}=\opu{A^{+}} \in  L(\hat{\cal K},\hat{\cal K})$ which 
accomplishes
\begin{equation}\label{(herma)}
\bd{x}\!A^{+}\gd\!\! \kd{y}=\bd{y}\!\!\gd A\!\kd{x}^{*}\,, \quad \forall 
\kd{x},\kd{y}\in {\cal K}\,. 
\end{equation}  
For $\opu{B}\in \linu$ the hermitian conjugation is defined by
\begin{equation}\label{(hermb)}
\bu{\hat x}\!B^{+}\gu\!\! \ku{\hat y}= 
\bu{\hat y}\!\!\gu B\!\ku{\hat x}^{*}\,,
\quad \forall \ku{\hat x},\ku{\hat y}\in \hat{\cal K}\,, 
\end{equation}  
where $\idd B^{+}\idd=(\idu B\idu)^{+}\in L({\cal K},{\cal K})$.
\end{defin}
One can convince ourselves that the mutually related bra vectors  with 
the ket vectors $/A\kd{x}$ and $\idu B\ku{y}$ are $\phi[/A\kd{x}]=
\bd{x}A^{+}\idu$ and $\hat \phi[\idu B\ku{y}]=\bu{y}B^{+}/$ respectively. 
\begin{defin}
We say that the operators which satisfy
\begin{equation}
\opu{A^{+}}=\gd A\gu \,, \quad \opd{B^{+}}=\gu B\gd\,,
\end{equation}
are hermitian with respect to the metric $\gd$, or simply {\em semi-hermitian}.
\end{defin}   
In other words an operator $/A/\in \lind$ is semi-hermitian if its adjoint 
operator coincides with its coupled operator $\idu \hat A\idu$ defined by 
(\ref{(asop)}). 
We specify that  these definitions can be formulated in terms of dual form.
For example, if we take $\ku{\hat y}=\gd\!\!\kd{y}$ then (\ref{(herma)}) and 
(\ref{(hermb)}) can be rewritten as
\begin{eqnarray}
\bd{x}\!A^{+}\!\ku{\hat y}&=&\bu{\hat y}\!A\!\kd{x}^{*}\,,\quad
\forall\, \kd{x}\in {\cal K},\,  \ku{\hat y}\in \hat{\cal K}\,,\label{(hermab)}\\
\bu{\hat x}\!B^{+}\!\kd{y}&=&\bd{y}\!B\!\ku{\hat x}^{*}\,,\quad 
\forall\, \kd{y}\in {\cal K},\,\ku{\hat x}\in \hat{\cal K}.  
\end{eqnarray}
Particularly, from (\ref{(bk1)}) we can draw the 
conclusion that the identity operators are semi-hermitian,  $(\Idd)^{+}=\Idu$. 

In practice it is convenient to introduce another conjugation operation 
which should unify the above definitions. This is just the generalization of 
the familiar Dirac conjugation of the theory of four-component spinors. 
\begin{defin}\label{(adjd)}
Given an operator $X$, the {\em Dirac adjoint} operator of $X$ is
\begin{equation}
\overline{X}=\left\{
\begin{array}{cll}
\gu X^{+}\gd& if &X\in \lind\,,\\              
\gd X^{+}\gu& if &X\in \linu\,,\\              
 X^{+}& if &X\in L({\cal K},\hat{\cal K})\, or\,  L(\hat{\cal K},{\cal K}) \,.              
\end{array}\right.
\end{equation}
The operator $X$ is {\em self-adjoint} if $\,\overline{X}=X$. 
\end{defin}
According to this definition all the hermitian and  semi-hermitian operators 
are self-adjoint, including  the metric and the identity operators,  
\begin{equation}
\overline{\gd}=\gd\,,\quad
\overline{\gu}=\gu\,,\quad
\overline{\Idd}=\Idd\,,\quad
\overline{\Idu}=\Idu\,.
\end{equation}
Furthermore, it is not difficult to demonstrate that 
\begin{equation}
\overline{\left(\overline{X}\right)}=X\,,    
\end{equation}
and that for two operators, $A$ and $B$,  from the same linear space $L$ and 
$\alpha, \beta \in \Comp$ we have
\begin{equation}
\overline{(\alpha A+\beta B)}=\alpha^{*}\,\overline{A}+\beta^{*}\,
\overline{B}\,.
\end{equation}
If the multiplication of two operators makes sense,  
e.g. $A,\,B\in \lind$ or $A\in L({\cal K},\hat{\cal K})$
and $B\in L(\hat{\cal K},{\cal K})$, etc, then we can show that
\begin{equation}
\overline{AB}=\overline{B}\,\overline{A}\,.
\end{equation}  
Thus we obtain a simple and homogeneous set of calculation rules for all the 
linear operators we manipulate. However, despite of this advantage, we prefer 
to use here the hermitian conjugation rather than the Dirac one since we are 
interested to follow the coherence of the presented formalism in its all 
details.

\section{Projection operators}

In the case of our cvs, where the orthogonality is defined by the metric 
operator, the problem of the decomposition in orthogonal subspaces is more 
complicated than that of unitary spaces \cite{ON} and, therefore, the theory of 
projection operators needs some specifications.   

Let us consider the cvs $({\cal K},\hat{\cal K},\gd)$ and an idempotent 
operator $/P/$, satisfying $P^{2}=P$, coupled with $\idu \hat P\idu =
\gd P\gu$ . Then  $P$ is the projection operator on the subspace 
$P{\cal K}\subset 
{\cal K}$ while  $\opu{\hat P}$, is the projection operator on the  subspace  
$\hat P\hat{\cal K}\subset \hat{\cal K}$. Two projection operators, 
$/P_{1}/$ and 
$/P_{2}/$, are {\em additive} if  $P_{1}P_{2}= P_{2}P_{1}=0$ since 
then $P_{1}+P_{2}$ is a projection operator  too. 
The operators $P_{a}, a=1,2,...,n$, which satisfy
\begin{equation}
P_{a}P_{b}=\delta_{a}^{b}P_{a}\,,   \quad \not\!\!{\Sigma}\,, 
\end{equation}
form  a set of additive projection operators. This set is called 
{\em complete} if 
\begin{equation}
\sum_{a}P_{a}=\Idd\,.
\end{equation}       
However, these projection operators  do not have good orthogonality 
properties since the additive ones generally are not orthogonal. For this 
reason one prefers the term perp instead of orthogonal \cite{ON}. 
\begin{defin} 
Two projection operators, $P_{1}$ and $P_{2}$, are perp if they satisfy
\begin{equation}
P_{1}^{+}\gd P_{2}=0
\end{equation}
Then the projection subspaces are called perp to each other.
\end{defin}

The projection subspaces of these projection operators can not be 
coupled anytime because there is the risk to find that the restriction of 
the metric operator to these subspaces is no more hermitian. Two 
projection subspaces,  $P{\cal K}$ and $\hat P\hat{\cal K}$, can be coupled 
only if their metric operator $\hat P\gd P$ is hermitian. This requires  
$\hat P=P^{+}$ which means that  $P$ must be semi-hermitian (self-adjoint). 
Then this has the property   
\begin{equation}\label{(vpr)}
P^{+}\gd =\gd P\,,  
\end{equation}
which indicates that the  subspaces $P{\cal K}$ and $P^{+}\hat{\cal K}$ are 
{\em invariant} subspaces of the metric operators $\gd$ and $\gu$. In these 
conditions the operators
\begin{equation}
\gd_{P}=P^{+}\gd P= P^{+}\gd = \gd P\,,\quad
\gu_{P}=P\gu P^{+}= P\gu = \gu P^{+},
\end{equation} 
can be considered as metric operators since they are hermitian and 
invertible in the sense that 
\begin{equation}
\gd_{P}\gu_{P}=P^{+}\,,\quad
\gu_{P}\gd_{P}=P\,.
\end{equation} 
\begin{defin}
The  invariant subspaces  $P{\cal K}$ and $P^{+}\hat{\cal K}$  isometric 
through $\gd_{P}$ represent a pair of coupled subspaces {\em (css)} denoted by  
$(P{\cal K},P^{+}\hat{\cal K}, \gd_{P}) 
\subset ({\cal K}, \hat{\cal K}, \gd)$.     
\end{defin}
The related css, $(\hat {\cal B}P, {\cal B}P^{+}, \gu_{P})$, can be 
defined using the restrictions to $P{\cal K}$ and 
$P^{+}\hat{\cal K}$ of the mappings $\phi$ and $\hat \phi$.     

Now the theory of orthogonal decomposition can be done in terms of css 
determined by semi-hermitian projection operators.
\begin{defin}
If two  semi-hermitian projection operators are perp then they as well as their 
projection css are called orthogonal. 
\end{defin}
This definition is justified by the fact that the semi-hermitian projection 
operators have similar properties as the usual hermitian ones. 
\begin{theor}
Two semi-hermitian projection operators are orthogonal if and only if they are 
additive.
\end{theor}
\begin{demo} Let us consider that  $P_{1}$ and $P_{2}$ are semi-hermitian and 
additive, satisfying $P_{1}P_{2}=0$. Then, according to (\ref{(vpr)}), we can 
write $0=\gd P_{1}P_{2}=P_{1}^{+}\gd P_{2}$ which means that these projection 
operators are orthogonal. From the same relation it results that the orthogonal 
projection operators must be additive since, by hypothesis, $\gd$ is invertible.
\end{demo}\\   
The consequence is that the css 
$({\cal K}_{3},\hat {\cal K}_{3}, \gd_{3})$ 
defined by the projection operator $P_{3}=P_{1}+P_{2}$ is the {\em direct sum} 
\begin{equation}   
({\cal K}_{3},\hat {\cal K}_{3}, \gd_{3})= 
({\cal K}_{1},\hat {\cal K}_{1}, \gd_{1})\oplus 
({\cal K}_{2},\hat {\cal K}_{2}, \gd_{2}) 
\end{equation}
where
\begin{equation}
\gd_{3}=\gd (P_{1}+P_{2})=\gd_{1}+\gd_{2}\,.
\end{equation}
If  $P_{2}=\Idd-P_{1}$ then the css defined by $P_{1}$ and $P_{2}$ are {\em 
orthogonal complements}. In general, a complete set of additive 
projection operators is  a complete set of orthogonal projection operators  if 
all of of these operators are  semi-hermitian.

\section{Matrix representations}

\subsection{Dual bases}

According to the usual terminology, any system of dual bases defines 
a {\em matrix representation} of the  cvs of ket and bra vectors. 
In our new notation the  vectors of the dual bases we have introduced in 
section 2 are 
\begin{eqnarray}
\kd{(i)}\equiv e_{i}\in {\cal K} \,, &\quad& 
\ku{(i)}\equiv \hat e^{i}\in \hat{\cal K}\,,\label{(b1)}\\
\bu{(i)}\equiv {\bar e}^{i} \in \hat{\cal B}\,, &\quad& 
\bd{(i)}\equiv \bar{\hat e}_{i}\in {\cal B}\,.\label{(b2)}
\end{eqnarray}
We consider that the indices $\idu(i)$ or $(i)/$ are in {\em upper} position 
while $/(i)$ or $(i)\idu$ are in {\em lower} position and we use the summation  
convention over dummy indices in opposite positions (e.g. we sum over $i$ in 
expressions where we find $... \idu (i)...(i)\idu...$ or $... \idd (i)...(i)
\idd...$). 

In a given representation, the main tools of our formalism are the 
duality conditions (\ref{(dual)}), written now as 
\begin{equation}\label{(dc)}
\bkud{(i)}{(j)}=\delta^{i}_{j}\,,\quad
\bkdu{(i)}{(j)}=\delta_{i}^{j}\,,
\end{equation}
and the {\em completeness relations}
\begin{equation}\label{(cr)}
\kd{(i)}\bu{(i)}=\Idd\,, \quad \ku{(i)}\bd{(i)}=\Idu\,,
\end{equation}
which show that  the sets of {\em elementary} projection operators
\begin{equation}\label{(epr)}
/P_{i}/=\kd{(i)}\bu{(i)} \quad \not\!\!{\Sigma}\,,\quad i=1,2,...,N,
\end{equation} 
and respectively $P_{i}^{+}, i=1,2,...,N$, are complete sets of additive
projection operators. 

These formulas contain all the information concerning the vector calculus 
with upper and lower indices, allowing us to express the final results  in 
terms of vector components and matrix elements. A special role play the 
matrix elements of  the  metric operator,  
\begin{eqnarray}
\bkdd{(i)}{(j)}&=&
\bd{(i)}\!\!\gd\!\!\kd{(j)}\equiv\eta_{ij}\,,\\
\bkuu{(i)}{(j)}&=&
\bu{(i)}\!\!\gu\!\!\ku{(j)}\equiv
{(\eta^{-1})}^{ij}\,,
\end{eqnarray}
which change the positions of indices. The components of ket-down vectors are  
$\bkud{(i)}{x}\equiv x^{i}$ or $\bkdd{(i)}{x}\equiv  x_{i}=\eta_{ij}x^{j}$, 
and similarly for the ket-up vectors. The components of the corresponding bra 
vectors have to be obtained through complex conjugation, according to the 
usual properties of the dual forms or scalar products. 

However, the most important is that now we are able to     
explicitly use spectral representations. For example, the operator 
$\opd{A}$ can be written as  
\begin{equation}
\opd{A}=\kd{(i)}\bu{(i)}\!A\!\kd{(j)}\bu{(j)}=\kd{(i)}A^{i\,\cdot}_{\cdot\,j}
\bu{(j)}
\end{equation}
where $A^{i\,\cdot}_{\cdot\,j}$ are the matrix elements in usual notation. 
The adjoint operator of $A$ is 
\begin{equation}
\opu{A^{+}}=\ku{(i)}\bd{(i)}\!A^{+}\!\ku{(j)}\bd{(j)}=\ku{(i)}
(A_{\cdot\,i}^{j\,\cdot})^{*}\bd{(j)}
\end{equation}\label{(hermaa)}
since from (\ref{(hermab)}) we have
\begin{equation}\label{(hermaa)}
(A^{+})_{i\,\cdot}^{\cdot\,j}\equiv \bd{(i)}\!A^{+}\!\ku{(j)}=
\bu{(j)}\!A\!\kd{(i)}^{*}\equiv(A_{\cdot\,i}^{j\,\cdot})^{*}\,.
\end{equation}
If, in addition, $A$ is semi-hermitian then 
$\bd{(i)}\!A^{+}\!\ku{(j)}=\bd{(i)}\!\!\gd A\gu\!\!\ku{(j)}$
and using (\ref{(hermaa)}) we recover the well-known property  
 $\eta_{ik}A^{k\,\cdot}_{\cdot\,l}(\eta^{-1})^{lj}=
(A_{\cdot\,i}^{j\,\cdot})^{*}$.

In any system of dual bases the traces of the operators $A\in \lind$ and 
$B\in \linu$ are defined by
\begin{equation}
{\rm Tr}(A)=\bu{(k)}\!A\!\kd{(k)}\equiv A^{k\,\cdot}_{\cdot \, k}\,,\quad
{\rm Tr}(B)=\bd{(k)}\!B\!\ku{(k)}\equiv B_{k\,\cdot}^{\cdot \, k}\,.
\end{equation}
It is easy to show that the coupled operators (\ref{(asop)}) have the same 
trace.

\subsection{Orthonormal bases}

In applications one prefers the  representations given by the {\em orthonormal} 
dual bases where the matrix of the metric operator is diagonal,   
\begin{equation}\label{(etad)}
\eta_{ij}\equiv\bkdd{(i)}{(j)}=\eta_{i}\delta_{ij} \quad \not\!\!{\Sigma}\,,\quad 
\eta_{i}=\pm1\,,
\end{equation} 
since there the ``squared norms" have the simplest expressions, 
\begin{equation}
\bkdd{x}{x}=\sum_{i=1}^{N}\eta_{i}|\bkud{(i)}{x}|^{2}\,.
\end{equation}  
The numbers $\eta_{i}$ may take $n_{+}$ times the value $1$ and $n_{-}=N-n_{+}$ 
times the value $-1$. Thus in orthonormal bases the metric operator is defined 
by its {\em signature} that can be given either explicitly as a sequence of 
signs or simply as $(n_{+},n_{-})$. We note that $n_{-}$ is called the 
{\em index} of the metric operator \cite{ON}.

The main advantage of the orthonormal bases is that there the elementary 
projection operators (\ref{(epr)}) are semi-hermitian and, consequently, they 
are orthogonal to each other. Of a particular interest are the orthogonal 
projection operators  
\begin{equation}
P_{+}=\sum_{\eta_{i}=1}P_{i}\,, \qquad 
P_{-}=\sum_{\eta_{i}=-1}P_{i}\,, 
\end{equation}    
which satisfy  $/P_{+}/+/P_{-}/=\Idd$.   
They split the space ${\cal K}$ into the pair of unitary spaces
${\cal K}_{+}=P_{+}{\cal K}$ and ${\cal K}_{-}=P_{-}{\cal K}$, of dimensions 
${\rm dim}\,{\cal K}_{+}=n_{+}$ and  ${\rm dim}\,{\cal K}_{-}=n_{-}$. Since 
the coupled space can be split in the same manner we have 
\begin{equation}\label{(deco)}
({\cal K},\hat{\cal K},\gd)=
({\cal K}_{+},\hat{\cal K}_{+},\gd_{+})\oplus
({\cal K}_{-},\hat{\cal K}_{-},\gd_{-})
\end{equation}
where
\begin{equation}
\gd=\gd_{+} +\gd_{-}\,;\quad \gd_{+}=\gd P_{+}\,,\quad  \gd_{-}=\gd P_{-}\,.
\end{equation}
Obviously, the same decomposition can be done for the related cvs of bra 
vectors. 

Finally we specify that in the particular case of the metric operators with 
signature  $(N,0)$ or $(0,N)$, 
the matrix $|\eta|$ in orthonormal bases coincides up to sign with the unit 
matrix.  Then our scalar products become  usual inner forms (i.e. scalar 
products in the sense of the theory of Hilbert spaces) and, therefore,  
${\cal K}$ and $\hat{\cal K}$ will be {\em unitary} spaces. In this 
situations we have two options. The first one is to keep the cvs structure if 
this is appropriate for our  problem. The second option is to consider 
only one ket space ${\cal K}\equiv \hat{\cal K}$ in usual Dirac formalism, 
with  $\ku{~}=\kd{~}=\ket{~}$ and $\gd=\gu=\Idd=\Idu=I$, where $I$ is the 
identity operator on ${\cal K}$.     

\section{Basis transformations}

\subsection{General linear transformations}

The change of the system of dual bases of our related pairs of cvs can be done  
using for each basis an arbitrary linear transformation, but then it is 
possible to obtain new bases 
which do not satisfy the canonical duality conditions or giving a non-hermitian 
matrix for the metric operator. In order to avoid these unwanted 
situations, we consider only the  transformations which preserve 
the cvs structure in the sense that (i) leave invariant the duality conditions 
and (ii) transform the hermitian matrix $|\eta|$ into another hermitian matrix, 
$|\eta'|$. 
\begin{theor}
The   general form of a transformation which satisfies the conditions (i) and 
(ii) is  
\begin{eqnarray}
&&\kd{(i)}\to\kd{(i)'}=\idd T\kd{(i)},\nonumber\\
&&\ku{(i)}\to\ku{(i)'}=\idu (T^{-1})^{+}\ku{(i)},\label{(cant)}\\
&&\bu{(i)}\to \bu{(i)'}=\bu{(i)}T^{-1}\idd,\nonumber\\
&&\bd{(i)}\to\bd{(i)'}=\bd{(i)}T^{+}\idu.\nonumber
\end{eqnarray} 
where $T\in {\rm Aut}\,({\cal K})\subset L({\cal K},{\cal K})$. 
\end{theor}
\begin{demo}
Let us start with the following  linear transformation of the ket 
vectors
\begin{equation}\label{(trket)}
\kd{(i)'}=\idd T\kd{(i)}\,,\quad \ku{(i)'}=\idu\tilde T\ku{(i)}\,,
\end{equation}
given by two  operators $T\in {\rm Aut}({\cal K})$ and $\tilde T\in {\rm Aut}
(\hat{\cal K})$ arbitrarily chosen. The transformations of the bra vectors 
which conserve the duality conditions are 
\begin{equation}\label{(trbra)}
\bu{(i)'}=\bu{(i)}T^{-1}\idd\,,\quad \bd{(i)'}=\bd{(i)}{\tilde T}^{-1}\idu\,.
\end{equation}
In these new bases the matrix of the metric operator remains hermitian only if 
$\tilde T=(T^{-1})^{+}$ since then we have  
\begin{equation}\label{(etap)}
\eta_{ij}'\equiv\bkdd{(i)'}{(j)'}=\bd{(i)}\! T^{+}\gd T\!\kd{(j)}=
(T^{k\,\cdot}_{\cdot\,i})^{*}T^{l\,\cdot}_{\cdot\,j}\,\eta_{kl}\,.
\end{equation}
Consequently, the  transformations (\ref{(trket)}) and (\ref{(trbra)}) take 
the form (\ref{(cant)}). 
\end{demo}

The conclusion is that $T$ is the operator of a {\em general linear} 
transformation of the  group $GL(N,\Comp)\subset \lind$. This means that 
our formalism is   equivalent  with the representation theory of 
general linear transformations  in the vector spaces of the theory of 
general tensors. Indeed, in a given system of dual bases, the components of the  
ket-down and respectively ket-up vectors transform according to a pair of
unequivalent fundamental representations of the $GL(N,\Comp)$ group while 
the bra components transform according to the corresponding complex conjugate  
representations. Moreover, we observe that  
the bra  components of our formalism, $\bkdu{x}{(k)}\equiv (x^{k})^{*}$ and 
$\bkud{\hat y}{(k)}\equiv (\hat y_{k})^{*}$, are just those replaced in the 
theory of general tensors  
by components carrying dotted indices in opposite positions. With this 
specification, it is clear that the four transformations laws (\ref{(cant)}) 
correspond to the four  unequivalent fundamental representations of the 
$GL(N,\Comp)$ group \cite{W}. 

The advantage of our formalism is that we work directly with the operators 
$T$  instead of the  matrices of their four representations. In this way we can     
easily find the transformation laws of the matrices of all the operators we  
use. The matrices of the operators $/A/$ and $\idu B\idu$ transform as 
\begin{eqnarray}
A^{i\,\cdot}_{\cdot\,j}\to
A^{'i\,\cdot}_{\cdot\,j}\equiv \bu{(i)'}\!A\!\kd{(j)'}&=&
\bu{(i)}\!T^{-1}AT\!\kd{(j)}\nonumber\\
&=&(T^{-1})^{i\,\cdot}_{\cdot\,k}
A_{\cdot\,l}^{k\,\cdot}T^{l\,\cdot}_{\cdot\,j} \\
B_{i\,\cdot}^{\cdot\,j}\to
B_{i\,\cdot}^{'\cdot\,j}\equiv \bd{(i)'}\!B\!\ku{(j)'}&=&
\bd{(i)}\!T^{+}B(T^{-1})^{+}\!\ku{(j)}\nonumber\\
&=&(T_{\cdot\,i}^{k\,\cdot})^{*}
B^{\cdot\,l}_{k\,\cdot}\left[(T^{-1})^{j\,\cdot}_{\cdot\,l}\right]^{*}
\end{eqnarray}
while those of the operators from $L({\cal K},\hat{\cal K})$ and 
$L(\hat{\cal K}, {\cal K})$ transform like the matrices of the metric 
operators, according to equation (\ref{(etap)}). 
These  transformations leave invariant the traces of the operators 
from $L({\cal K},{\cal K})$ and $L(\hat{\cal K},\hat{\cal K})$. Moreover, 
one can show that any system of dual bases can be transformed at any time in 
a system of othonormal bases using a suitable general transformation.

\subsection{Symmetry transformations}

The general transformations  change the form of all the operator 
matrices including that of the metric operator. However, there is a special 
case of some  transformations which  do not change the matrix of the 
metric operator.   
\begin{defin}
The general transformations that  leave invariant the matrix $|\eta|$ 
are called symmetry transformations.  
\end{defin}
These  transformations  are of the form (\ref{(cant)}) but their operators 
have special properties.  
\begin{theor}
The operators $U$ of the symmetry transformations must satisfy the 
{\em semi-unitarity} condition, 
\begin{equation}\label{(etaun)}
\idu U^{+}\gd U\idd=\gd\,,
\end{equation}
which can be written as  $\overline{U}=U^{-1}$.
\end{theor}  
\begin{demo}
A transformation (\ref{(cant)}) generally change the matrix $|\eta|$ 
according to (\ref{(etap)}). This is a symmetry transformation only 
if $\bkdd{(i)'}{(j)'}=\bkdd{(i)}{(j)}$. Hereby it results (\ref{(etaun)}).
\end{demo}\\ 
Consequently, the symmetry transformations have the form
\begin{eqnarray}
&&\kd{(i)}\to\kd{(i)'}=\idd U\kd{(i)},\nonumber\\
&&\ku{(i)}\to\ku{(i)'}=\idu \hat U\ku{(i)},\label{(cantu)}\\
&&\bu{(i)}\to \bu{(i)'}=\bu{(i)} U^{-1}\idd,\nonumber\\
&&\bd{(i)}\to\bd{(i)'}=\bd{(i)}\hat U^{-1}\idu.\nonumber
\end{eqnarray} 
where we recall that $\hat U=\gd U\gu$ is the operator coupled with $U$.
The main virtue of these transformations is that they do not change the 
expressions of scalar product in terms of vector components. In other words 
these leave invariant not only the expressions of the dual forms but also those of 
the scalar products. Moreover, when we work with orthonormal bases the symmetry 
transformations conserve the orthogonality.

The  semi-unitary operators $U$ which accomplish the  condition 
(\ref{(etaun)}) form a subgroup of $GL(N,\Comp)$, namely the {\em maximal 
symmetry group} or the {\em gauge group} of the 
metric operator $\gd$. Any system of dual bases defines a pair of 
{\em coupled} fundamental representations of this group and its algebra 
in the carrier spaces ${\cal K}$ and $\hat{\cal K}$. 
Obviously, these representations are {\em equivalent} through the metric 
operator.  

A given  pair of ket cvs can be used as carrier spaces for the 
coupled {\em semi-unitary}  representations 
of any subgroup of the gauge group. In general, these 
representations are  reducible in  usual sense but it is not sure that their 
subspaces can be correctly coupled. For this reason we consider here a modified 
definition of reducibility.   
\begin{defin}\label{(redu)}
The  coupled representations are  irreducible 
if their generators and the metric operator have no common non-trivial 
invariant subspaces. Otherwise the representations are reducible.
\end{defin}
Since these representations are semi-unitary one can show that, 
like in the unitary case, the reducible representations are decomposable (i.e. 
fully reducible). Consequently, the original cvs can be written as a direct sum 
of css carrying irreducible representations. Thus, using this definition we 
preserve  the  important advantage of the css structure that  guarantees the 
existence of invariant scalar products. 

\subsection{Group generators}

The properties of the $GL(N,\Comp)$ group and its subgroups  are 
well-studied but it is interesting to review few among them  in  our 
formalism where we can work directly with the spectral representations of the 
group generators. 

Let us consider the  related pairs of cvs  carrying the  fundamental 
representations of the $GL(N,\Comp)$ group (\ref{(cant)}). In any system of 
dual bases (\ref{(b1)}) and (\ref{(b2)}) the usual  parametrisation of the 
operators $T\in GL(N,\Comp)\subset \lind$ reads  
\begin{equation}\label{(param)}
T(\omega)=e^{\omega^{ij}X_{ij}}\,, \quad X_{ij}=\kd{(i)}\bd{(j)}\!\!\gd,
\end{equation}
where  $\omega^{ij}$ are arbitrary c-numbers and $X_{ij}$ are the ``real 
generators" (having  real matrix elements in orthonormal bases) which satisfy 
\begin{equation}\label{(hermx)}
(X_{ij})^{+}=\gd X_{ji}\gu \qquad \left(\overline{X}_{ij}=X_{ji}\right)\,.
\end{equation}
Hereby we can separate the $SL(N,\Comp)$ generators 
\begin{equation}\label{(hij)}
H_{ij}=X_{ij}-\frac{1}{N}\eta_{ji}\,\Idd\,,
\end{equation}
which have the properties
\begin{equation}
\bkuu{(i)}{(j)}H_{ij}=0\,,\quad {\rm Tr}(H_{ij})=0\,.
\end{equation}

If we consider only the gauge group of the metric operator $\gd$ then we 
have to use the same generators but with restrictions imposed upon 
the  parameter values. From  equations (\ref{(etaun)}) and (\ref{(hermx)}) 
we obtain that the parameters of the gauge group must satisfy 
the condition $\omega^{ij}+(\omega^{ji})^{*}=0$, which means that
\begin{equation}\label{(roio)}
\Re\, \omega^{ij}=-\Re\, \omega^{ji}\,,\quad \Im\,\omega^{ij}=
\Im\,\omega^{ji}\,.
\end{equation}
There are $N^{2}$ real parameters  as it was expected since  the gauge group of 
the metric  operator $\gd$  of signature  $(n_{+},n_{-})$  is the 
{\em semi-unitary} group $U(n_{+},n_{-})=U(1)\otimes SU(n_{+},n_{-})$. 
When one uses the parameters $\omega ^{ij}$ then the operators (\ref{(hij)}) 
are considered as  ``real" $SU(n_{+},n_{-})$ generators. However, the 
canonical parametrisation with the {\em real} parameters (\ref{(roio)}) 
reads $\omega^{ij}H_{ij}=-i(\Re\omega^{ij}A_{ij}+\Im\omega^{ij}S_{ij})$ 
involving  the {\em semi-hermitian} (self-adjoint) $SU(n_{+},n_{-})$ 
generators defined as
\begin{eqnarray}
A_{ij}&=&\frac{i}{2}(X_{ij}-X_{ji})+\frac{1}{N}\Im\,\eta_{ji}\,\Idd\,,\nonumber \\
S_{ij}&=&-\frac{1}{2}(X_{ij}+X_{ji})+\frac{1}{N}\Re\,\eta_{ji}\,\Idd\,.\label{(AS)}
\end{eqnarray}
It is clear that the antisymmetric ones, $A_{ij}$,  are the generators of 
the subgroup $SO(n_{+},n_{-})\subset SU(n_{+},n_{-})$ \cite{BR}. 
In the particular case of real cvs the gauge group reduces 
to $O(n_{+},n_{-})$. On the other hand, with the help of the generators 
(\ref{(AS)}) one can show that the group $SL(N,\Comp)$ is the complexification 
of the $SU(n_{+},n_{-})$ group in the sense that in a parametrisation with 
real numbers the $SL(N,\Comp)$ generators are $A_{ij}$, $S_{ij}$, $iA_{ij}$ 
and $iS_{ij}$.

An interesting problem is how transform the generators $X_{ij}$ 
when we change the bases  through a general linear transformation  (\ref{(cant)}).  
In our formalism it is easy to show that, according to (\ref{(param)}), the 
transformed generators are  
\begin{equation}
X'_{ij}=\kd{(i)'}\bd{(j)'}\!\!\gd=T X_{ij}\gu T^{+}\gd=TX_{ij}\overline{T}.
\end{equation}
Particularly, if we consider only  symmetry transformations, $T=U$, 
then from (\ref{(etaun)}) we recover the usual transformation law
of the $SU(n_{+},n_{-})$ generators, 
\begin{equation}
H'_{ij}=U H_{ij}U^{-1}=U^{k\,\cdot}_{\cdot\,i}
(U^{l\,\cdot}_{\cdot\,j})^{*}H_{kl}\,,
\end{equation} 
which indicates that they  transform according to the {\em adjoint} 
representation of  $SU(n_{+},n_{-})$.

\section{The semi-unitary representations of the $SL(2,\Comp)$ group}

The application presented in order to illustrate how works our formalism is 
the problem of the finite-dimensional  representations of the $SL(2,\Comp)$ 
group with invariant scalar products. It is well-known that the 
finite-dimensional irreducible representations of the $sl(2,\Comp)$ algebra  
can be constructed with the help of those of the $su(2)$
algebra \cite{W}. However, in general, these do not have  invariant scalar
products under  $SL(2,\Comp)$ transformations. In practice these scalar
products are defined in each
particular case of physical interest separately starting with a suitable
representation which is often reducible. In this section we would like to
present the general theory of the semi-unitary and irreducible
finite-dimensional representations of $sl(2,\Comp)$, in coupled carrier 
spaces where the invariant scalar products are well-defined.          

\subsection{The representations of $su(2)$ in cvs}

The problem of the irreducible representations of $su(2)$ in cvs reduces to 
that of the canonical irreducible representations in unitary spaces 
${\cal K}^{j}\sim \hat{\cal K}^{j}$ of weight $j$ \cite{AB}. Therefore the 
non-trivial 
cvs may be carrier spaces  only for reducible coupled representations. These cvs 
have the general structure   
\begin{equation}\label{(cvsrr)}
({\cal K},\hat {\cal K}, \gd)=\sum_{j\in {\bf J}}\oplus  
({\cal K}^{j},\hat {\cal K}^{j}, \gd^{j})  
\end{equation}
where 
\begin{equation}
\gd=\sum_{j\in {\bf J}} \gd^{j}
\end{equation}
and ${\bf J}$ is an arbitrary set of weights. In each css we consider the 
system of {\em canonical} bases of ket vectors, $\{\kd{j,\lambda}\}\subset 
{\cal K}^{j}$ and $\{\ku{j,\lambda}\}\subset \hat{\cal K}^{j}$,
$\lambda=-j,-j+1,...,j$, and the related bases of bra vectors satisfying  
\begin{equation}
\bkud{j,\lambda}{j',\lambda'}=\delta^{j}_{j'}\delta^{\lambda}_{\lambda'}\,,
\quad
\bkdu{j,\lambda}{j',\lambda'}=\delta_{j}^{j'}\delta_{\lambda}^{\lambda'}\,.
\end{equation}
The metric operators of css, 
\begin{equation}
\gd^{j}=\epsilon_{j}\sum_{\lambda}\ku{j,\lambda}\bu{j,\lambda}\,,
\end{equation}
are defined by the set of numbers $\epsilon_{j}=\pm 1$, $j\in {\bf J}$, that
gives the signature of the whole metric operator $\gd$ of the cvs 
(\ref{(cvsrr)}). For each css the metric operator $\gd^{j}$ couples the 
subspaces ${\cal K}^{j}=P^{j}{\cal K}$ and 
$\hat{\cal K}^{j}={P^{j}}^{+}\hat{\cal K}$ given by the semi-hermitian 
projection operator 
\begin{equation}
P^{j}=\sum_{\lambda}\kd{j,\lambda}\bu{j,\lambda}.
\end{equation}
The spectral representations of the  projections of the 
operators $X\in su(2)$ are
\begin{equation}
X^{j}=P^{j}XP^{j}=\sum_{\lambda\lambda'}\kd{j,\lambda}X^{j}_{\lambda,
\lambda'}\bu{j,\lambda'},
\end{equation}
where $X^{j}_{\lambda,\lambda'}$ are the usual matrix elements of $X$ in the 
canonical basis of the unitary irreducible representation of weight $j$.    
Hereby it is easily to verify that  the generators of the irreducible 
representations with values in $L({\cal K}^{j}, {\cal K}^{j})$, denoted by 
$/J^{j}_{a}/$, $a=1,2,3$,  are semi-hermitian. Consequently, the generators 
of the coupled representation are $\hat J_{a}^{j}=(J_{a}^{j})^{+} 
\in L(\hat{\cal K}^{j}, \hat{\cal K}^{j})$.  

\subsection{Finite-dimensional representations of $sl(2,\Comp)$}

The usual non-covariant generators of the  $sl(2,\Comp)$ algebra are the 
rotation generators, $I_{a},\, a=1,2,3$, and  the Lorentz boosts, 
$K_{a}$. From their well-known commutation rules it results that the algebras 
generated by 
\begin{equation}
M_{a}=\frac{1}{2}\left(I_{a}+iK_{a}\right)\,,\quad  
N_{a}=\frac{1}{2}\left(I_{a}-iK_{a}\right) 
\end{equation}
are two $su(2)$ algebras commuting with each other \cite{W}. Notice that 
these algebras can not be seen as ideals of $sl(2,\Comp)$ since their 
generators are complex linear combinations of the generators of a real 
algebra.

Our aim is to construct the semi-unitary finite-dimensional representations   
of $sl(2,\Comp)$ using our formalism. This means to consider from the 
beginning that the generators of the coupled representations  are 
semi-hermitian satisfying  $I_{a}^{+}=\gd I_{a}\gu=\hat I_{a}$ and 
$K_{a}^{+}=\gd K_{a}\gu=\hat K_{a}$ (or $\overline{I_{a}}=I_{a}$ and 
$\overline{K_{a}}=K_{a}$). Consequently, we must have  
\begin{equation}\label{(mnv)}
M_{a}^{+}=\gd N_{a}\gu=\hat N_{a}\,,\quad   
N_{a}^{+}=\gd M_{a}\gu=\hat M_{a}\,,\quad (\overline{M_{a}}=N_{a})\,.    
\end{equation}

The general solution of this problem can be written starting with the space 
of the reducible representation $(j_{1},j_{2})\oplus (j_{2},j_{1})$ of the 
$sl(2,\Comp)$ algebra \cite{W}, ${\cal K}=({\cal K}^{j_{1}}\otimes 
{\cal K}^{j_{2}})\oplus ({\cal K}^{j_{2}}\otimes {\cal K}^{j_{1}})$, with 
$j_{1}\not=j_{2}$. 
We find that the generators 
\begin{eqnarray}
M_{a}&=&J_{a}^{j_{1}}\otimes P^{j_{2}} + J_{a}^{j_{2}}\otimes P^{j_{1}}\,,\\   
N_{a}&=&P^{j_{1}}\otimes J_{a}^{j_{2}} + P^{j_{2}}\otimes J_{a}^{j_{1}}\,,   
\end{eqnarray}
and the metric operator 
\begin{eqnarray}
\gd=\epsilon_{j_{1},j_{2}}\sum_{\lambda\lambda'}\left(\ku{j_{1},\lambda}
\otimes \ku{j_{2},\lambda'}\bu{j_{2},\lambda'}\otimes \bu{j_{1},\lambda}
\right.\nonumber\\
\left.+ \ku{j_{2},\lambda'}\otimes
\ku{j_{1},\lambda}\bu{j_{1},\lambda}\otimes \bu{j_{2},\lambda'}\right)
\end{eqnarray}
with $\epsilon_{j_{1}j_{2}}=\pm 1$, satisfy equations (\ref{(mnv)}). 
Hereby  it results that the vector space coupled with ${\cal K}$ must be 
$\hat {\cal K}=({\cal K}^{j_{2}}\otimes {\cal K}^{j_{1}})
\oplus ({\cal K}^{j_{1}}\otimes {\cal K}^{j_{2}})$. 

These semi-unitary representations in cvs $({\cal K},\hat{\cal K},\gd)$ 
will be denoted by $[j_{1},j_{2}]$. They can be considered  
{\em irreducible} in the sense of definition (\ref{(redu)}) since the metric 
operator $\gd$ and the generators $M_{a}$ and $N_{a}$ do not have common 
non-trivial invariant subspaces. 
We can convince that with the help of the {\em chiral} projection operators, 
$/P_{L}/= P^{j_{1}}\otimes P^{j_{2}}$ and $/P_{R}/=P^{j_{2}}\otimes P^{j_{1}}$,
which generalize the familiar left and right-handed ones of the theory of 
Dirac spinors. These projection operators are just those of the invariant 
subspaces of the generators $M_{a}$ and $N_{a}$. They form a complete set 
of additive projection operators, 
\begin{equation}
P_{L}P_{R}=0\,,\quad P_{L}+P_{R}=\Idd,
\end{equation}
but they are not semi-hermitian (or self-adjoint) operators since
\begin{equation}
{P_{L}}^{+}=\gd P_{R}\gu\,,\quad    
{P_{R}}^{+}=\gd P_{L}\gu \,, \quad (\overline{P_{L}}=P_{R})\,.    
\end{equation}
Therefore, the subspaces $P_{L}{\cal K}$ and $P_{R}{\cal K}$ are not 
invariant subspaces of $\gd$ and the coupled representations $[j_{1},j_{2}]$ 
are irreducible from our point of view.   

In the particular case of $j_{1}=j_{2}=j$ the solution is simpler.   
The space ${\cal K}=\hat{\cal K}={\cal K}^{j}\otimes {\cal K}^{j}$ is just 
that of the irreducible representation $(j,j)$ of $sl(2,\Comp)$ \cite{W}. 
The generators have the form 
\begin{eqnarray}
M_{a}&=&J_{a}^{j}\otimes P^{j}\\   
N_{a}&=&P^{j}\otimes J_{a}^{j}   
\end{eqnarray}
while the metric operator reads
\begin{equation}
\gd=\epsilon_{j}\sum_{\lambda\lambda'}\ku{j,\lambda}\otimes 
\ku{j,\lambda'}\bu{j,\lambda'}\otimes \bu{j,\lambda}\,.
\end{equation}
These irreducible representations will be denoted by $[j]$.

\subsection{Rotation bases}

Now we can introduce the system of rotation dual bases in which the operators 
$I^{2}=(I_{1})^{2}+(I_{2})^{2}+(I_{3})^{2}$ and $I_{3}$ as well as their 
coupling partners  are diagonal. Since $I_{a}=M_{a}+N_{a}$, the vectors of 
these 
bases  can be constructed with the help of the Clebsh-Gordan coefficients of 
the $SU(2)$ group \cite{AB}. The ket-down vectors of the rotation 
basis of the subspace ${\cal K}^{j_{1}}\otimes {\cal K}^{j_{2}}\subset 
{\cal K}$ are defined as   
\begin{equation}
\kd{(j_{1},j_{2})\,s,\sigma}=\sum_{\lambda,\lambda'=\sigma-\lambda}\kd{j_{1},\lambda}\otimes
\kd{j_{2},\lambda'}\,\bk{j_{1},\lambda;j_{2},\lambda'}{s,\sigma}
\end{equation}
such that
\begin{eqnarray}
I^{2}\kd{(j_{1},j_{2})\,s,\sigma}&=&s(s+1)\kd{(j_{1},j_{2})\,s,\sigma}\,,\\
I_{3}\kd{(j_{1},j_{2})\,s,\sigma}&=&\sigma\kd{(j_{1},j_{2})\,s,\sigma}\,.
\end{eqnarray}
The rotation bases of the other spaces of our cvs have to be introduced in 
the same manner. Then by taking into account that \cite{AB}
\begin{equation}
\bk{j_{1},\lambda;j_{2},\lambda'}{s,\sigma}=
(-1)^{s-j_{1}-j_{2}}\bk{j_{2},\lambda';j_{1},\lambda}{s,\sigma}
\end{equation}
and  using the  orthogonality  relations of these coefficients, we  
find the final forms of the metric operators in rotation bases. When 
$j_{1}\not=j_{2}$ this is
\begin{eqnarray}
\gd=\epsilon_{j_{1},j_{2}}(-1)^{-j_{1}-j_{2}}\sum_{s=|j_{1}-j_{2}|}
^{j_{1}+j_{2}}(-1)^{s}\sum_{\sigma=-s}^{s}(
\ku{(j_{1},j_{2})\,s,\sigma}\bu{(j_{2},j_{1})\,s,\sigma}\nonumber\\
+\ku{(j_{2},j_{1})\,s,\sigma}\bu{(j_{1},j_{2})\,s,\sigma})\,,
\end{eqnarray}
while for $j_{1}=j_{2}=j$ we have
\begin{equation}\label{(omj)}
\gd=\epsilon_{j}(-1)^{2j}\sum_{s=0}^{2j}(-1)^{s}\sum_{\sigma=-s}^{s}
\ku{(j,j)\,s,\sigma} \bu{(j,j)\,s,\sigma}\,. 
\end{equation}
In our opinion a good choice of the factors $\epsilon$ could be
\begin{equation}\label{(eee)}
\epsilon_{j_{1},j_{2}}=(-1)^{j_{1}+j_{2}-|j_{1}-j_{2}|}\,,\quad
\epsilon_{j}=(-1)^{2j}\,.
\end{equation}

Thus we obtain the spectral representations of the  metric operators in 
rotation dual bases. We observe that these bases are orthonormal  
only for $j_{1}=j_{2}$. In the general case of $j_{1}\not=j_{2}$ the 
ket-down vectors of the orthonormal basis of ${\cal K}$ are given by 
the linear combinations
\begin{equation}
\kd{(\pm)\,s,\sigma}=\frac{1}{\sqrt{2}}(\kd{(j_{1},j_{2})\,s,\sigma}\pm
\kd{(j_{2},j_{1})\,s,\sigma})\,.
\end{equation}
Similarly we get the  ket or bra vectors of the other orthonormal bases where  
the metric operator can be represented as
\begin{eqnarray}
\gd=\epsilon_{j_{1},j_{2}}(-1)^{-j_{1}-j_{2}}\sum_{s=|j_{1}-j_{2}|}
^{j_{1}+j_{2}}(-1)^{s}\sum_{\sigma=-s}^{s}(
\ku{(+)\,s,\sigma}\bu{(+)\,s,\sigma}\nonumber\\
-\ku{(-)\,s,\sigma}\bu{(-)\,s,\sigma})\,.
\end{eqnarray}
From this formula we see that for $j_{1}\not=j_{2}$ the metric operator has 
the symmetric signature $(n,n)$ with $n=(2j_{1}+1)(2j_{2}+1)$. The signatures 
of the metric operators of the representations $[j]$ result from (\ref{(omj)}) 
and (\ref{(eee)}) to be either $(n,m)$ for integer $j$ or $(m,n)$ if $j$ is a 
half-integer, where  $n=(j+1)(2j+1)$ and $m=j(2j+1)$.

Hence for each pair of coupled irreducible representations of $sl(2,\Comp)$ 
the metric operator has a well-determined  signature, $(n_{+},n_{-})$, which 
shows us that the corresponding maximal symmetry group is just $U(n_{+},n_{-})$. 
This result can be important from the physical point of view since  the 
transformations of this group are semi-unitary, leaving invariant the scalar 
product of the carrier spaces of the $sl(2,\Comp)$ representations.    
For example, in the theory of  Dirac spinors the metric operator of the 
representation $[1/2,0]$ has the signature $(2,2)$ which explains why the 
whole algebra of  $\gamma$ matrices and $sl(2,\Comp)$ generators is just 
the $u(2,2)$ algebra \cite{TY}. We note that the corresponding $U(2,2)$ group 
was recently considered as an extended gauge group of the theory of the Dirac 
field in curved spacetime, obtaining thus new interesting results \cite{F}.

\section{Concluding remarks}

The  presented formalism is the natural generalization of the Dirac's 
bra-ket calculus to spaces with indefinite metric. The main point of our  
proposal is to organize the four different spaces of the theory of  general 
tensors into a pair of coupled ket spaces mutually related with a pair of 
coupled bra one. Then the metric operator can be correctly introduced in     
accordance with the natural duality such that compatible scalar products and 
dual forms do co-exist. In this way we recover the results of the  theory of 
the general linear transformations formulated simply in terms of operators, 
independent on the concrete representations given by systems of dual bases.    
For this reason, our approach helps one to precise the nature of the 
different mathematical objects,  avoiding the risk to confuse 
among themselves those which accidently could have the same components or 
matrix elements in several representations (e.g. the components of second 
rank tensors and the operator matrix elements).  Moreover, in this framework 
some special operations used up to now only in particular problems get a
general meaning. We refer especially to the Dirac adjoint which in our 
formalism can be defined for all the types of involved operators, taking 
over the role of the Hermitian adjoint from the usual unitary case.

The presented example points out few of these advantages. Using our 
general bra-ket calculus we are able  to write the spectral representations 
in different basis of the $sl(2,\Comp)$ generators and the metric operators of 
the finite-dimensional irreducible representations with invariant scalar 
products. The study of basis transformations, the generalization of the 
chiral (left and right-handed) projection operators and the analyze of the
gauge group of the metric operator can be easily done in this context.
  
We tried here to remain in the spirit of the original 
Dirac's bra-ket calculus, imagining a  formalism with simple calculation 
rules but having the ``memory" of the definitions and properties of its basic 
elements. We hope that this should be appropriate for different applications 
including algebraic programming on computers.      

\subsection*{Acknowledgments} 

I would like to thank Mircea Bundaru and Lucian Gligor for useful discussions 
concerning the construction of the proposed formalism.

\end{document}